# Power spectral density estimation for wireless fluctuation enhanced gas sensor nodes

R. MINGESZ, G. VADAI, and Z. GINGL

*Department of Technical Informatics*
*University of Szeged, Árpád tér 2., Szeged, H-6720, Hungary*
*mingesz@inf.u-szeged.hu*

Fluctuation enhanced sensing (FES) is a promising method to improve the selectivity and sensitivity of semiconductor and nanotechnology gas sensors. Most measurement setups include high cost signal conditioning and data acquisition units as well as intensive data processing. However, there are attempts to reduce the cost and energy consumption of the hardware and to find efficient processing methods for low cost wireless solutions. In our paper we propose highly efficient signal processing methods to analyze the power spectral density of fluctuations. These support the development of ultra-low-power intelligent fluctuation enhanced wireless sensor nodes while several further applications are also possible.

*Keywords:* fluctuation enhanced sensing; wireless sensor network; spectral estimation.

# Introduction

Fluctuation enhanced sensing (FES) has become an active research area in the field of gas sensors since its introduction a few years ago [1]. Typically, measuring the DC value of sensor resistance is used to determine the gas concentration. In this case, each sensor is calibrated and prepared for a certain type of gas, therefore several sensors are used in parallel to detect and measure gas mixtures. The FES principle uses the time-dependent fluctuations of the sensors' resistance as an information source. Measuring these fluctuations caused by adsorption-desorption and diffusion noise provides enhanced selectivity and sensitivity, therefore it could be used to measure different gases or mixtures even with a single sensor. There are several examples of successful applications, including gas mixture detection, harmful gas and bacterial odor sensing using Taguchi sensors [2-4], increasing selectivity of nanotechnology gas sensors [5-9] and detecting scents using semiconductor sensors [10].

A typical measurement setup requires high cost signal conditioning and data acquisition units. Data processing is usually done offline by using high performance personal computers that perform complex calculations including spectral analysis, removal of external interferences and pattern recognition [5-7]. During the last few years, suggestions have been made to replace these complex systems by small, wireless data acquisition modules [11-13]. In addition, different methods with low processing needs have been simultaneously proposed [14, 15]. In the case of the binary fingerprint method [14], the power spectrum of the noise is divided into different regions. The signal power present in these regions is used to determine the components of a gas mixture. Different, band-pass and low-pass filter based units were designed to support this analysis method [16, 17].

Recently, we have developed a complete, standalone intelligent FES sensor node [13] based on the principle of measuring the variance of the noise at the outputs of eight first order low-pass filters with logarithmically distributed corner frequencies. These filters are used to estimate the power of the noise in different frequency regions [14]. In our current paper we propose two different power spectral density reconstruction methods using the same system and analyze their performance in detail. We also compare the performance of the proposed methods with Fourier-transformation based ones. Our measurement and analysis system contains the entire analogue and digital signal processing required for efficient universal spectral estimation in wireless sensing and sensor networks. Consequently, it has applications in several interdisciplinary fields other than FES as well.

# System

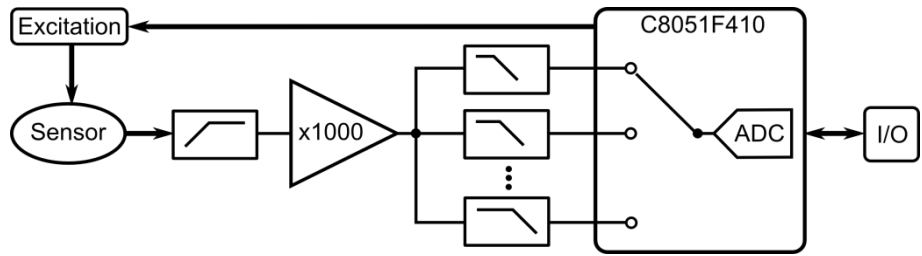

Fig. 1. Block diagram of the low power FES device

Fig. 1 contains the block diagram of the intelligent sensor node. The processing unit is a low-power high-performance mixed signal microcontroller (C8051F410). The unit is designed to process the noise of semiconductors (Taguchi) or nanotechnology based sensors. The excitation of the sensor is provided by a current generator with selectable current levels. A 1 Hz high-pass filter is used to remove the DC component of the sensor output, and an amplification of 1000 is applied to ensure proper signal levels. This signal is fed into a bank of first order passive RC low-pass filters. We estimated the power spectra of the noise by measuring only the average output power level of each filter. We used eight filters with the following logarithmically distributed nominal corner frequencies: 10 Hz, 27 Hz, 72 Hz, 193 Hz, 518 Hz, 1389 Hz, 3728 Hz and 10000 Hz. The outputs of the filters are selected using a built-in analogue multiplexer of the microcontroller and then digitized by a precision 12-bit analogue-to-digital converter. An aggregate sampling frequency of 8 kHz was used, resulting a 1 kHz effective sampling frequency for each filter. Note that while this frequency does not satisfy the sampling theorem, we need to measure the output power level of the output only, which can be measured even by undersampling, since the signal and the sampling time instants are uncorrelated. The output variance of each filter can be sent to a host computer via USB or wireless link for further processing. However, our aim was to find efficient signal processing algorithms for recognizing gas concentration and gas mixtures directly by the sensor node itself. The power consumption of the system, the effect of the precision of the filters and the effect of measurement length are evaluated in our previous publication [13].

# Methods

Our proposed data processing method is based on the principle of approximating the resistance fluctuation's power spectral density (PSD) using the measured power of the adjacent frequency bands at the outputs of the low-pass filter bank. We have examined the sensitivity and selectivity of the sensor node and evaluated the PSD reconstruction methods theoretically and by numerical simulations as well.

## Estimation of each filter's output power

We estimated the output power (variance) of each filter for signals with different power spectral densities. In order to calculate the variance, one can use the following equation:

$$\sigma_i^2 = \int_0^{f_{max}} \frac{\left(\frac{f}{f_h}\right)^2}{1+\left(\frac{f}{f_h}\right)^2} \cdot \frac{1}{1+\left(\frac{f}{f_i}\right)^2} \cdot X(f) \cdot df, \qquad (1)$$

where $X(f)$ is the PSD of the signal's output noise, $\sigma_i^2$ is the variance of $i^{th}$ filter's output, and $f_{max}$ is the bandwidth of the system. Since we wanted our method to work for arbitrary spectral dependence, we used numerical integration to calculate the result. The parameters of the integration are: $\Delta f$ = 0.1 Hz, $f_{max}$ = 1 MHz.

We also used numerically simulated measurements in order to approximate the output variance of the filter. By performing multiple measurements, we could also approximate the standard deviation (SD) of each measured variance. In order to be able to work with arbitrary spectral densities, we used Fourier-transformation based methods to generate an arbitrary signal as well as to implement low-pass and high-pass filters. Each signal was generated using a sampling frequency of 100 kHz, the duration was 10 s [13]. The output of each filter was resampled at 1 kHz to simulate the operation of the sensor node. Each variance value was calculated using 10000 samples.

## PSD reconstruction methods

The direct method for spectrum reconstruction assumes ideal low-pass and high-pass filters, thus power present in the $i$th frequency band is equal to $\sigma_i^2 - \sigma_{i-1}^2$. Based on this assumption the reconstructed PSD amplitude is given by:

$$y_{s_i} = \frac{\sigma_i^2 - \sigma_{i-1}^2}{f_i - f_{i-1}} \text{ if } i > 1, \text{ otherwise } y_{s_1} = \frac{\sigma_1^2}{f_1 - f_h}, \qquad (2)$$

where $f_i$ is the cut-off frequency of the $i$th filter and $f_h$ is the cut-off frequency of the high-pass filter. The x coordinates of the spectrum are given by the geometric mean of the beginning and the end of selected band:

$$f_i^* = \sqrt{f_i \cdot f_{i-1}} \text{ if } i > 1, \text{ otherwise } f_1^* = \sqrt{f_1 \cdot f_h}, \qquad (3)$$

While implementing this method is really simple, it does not take the transfer function of the applied filters into account. For this reason, we used an alternative method, the 1/$f$ *normalized method*.

Assuming, that we have a 1/$f$ noise at the output of the sensor, we can calculate the expected variances at the output of each filter according to the following formula:

$$\sigma_{Ni}^2 = \int_0^{f_{max}} \frac{\left(\frac{f}{f_h}\right)^2}{1+\left(\frac{f}{f_h}\right)^2} \cdot \frac{1}{1+\left(\frac{f}{f_i}\right)^2} \cdot \frac{1 \cdot V^2 Hz}{f \cdot f_{max}} \cdot df, \qquad (4)$$

Note that this noise is generated by 1/$f$ filtering of a white noise with variance of 1 V² and bandwidth of $f_{max}$. As in the case of direct method, we can also calculate the power of the $i$th frequency band: $\sigma_{Ni}^2 - \sigma_{Ni-1}^2$. Since the PSD of the measured noise will differ from the reference noise, the measured power in this band will be different from this value. The ratio between the power levels will be approximately equal to the ratio between the two PSD-s in this band. According to the previous assumption, we give the PSD approximation by the following formula:

$$y_{p_i} = \frac{\sigma_i^2 - \sigma_{i-1}^2}{\sigma_{Ni}^2 - \sigma_{Ni-1}^2} \cdot \frac{1 \cdot V^2 Hz}{f_i^* \cdot f_{max}} \text{ if } i > 1, \text{ otherwise } y_{p_1} = \frac{\sigma_i^2}{\sigma_{Ni}^2} \cdot \frac{1 \cdot V^2 Hz}{f_i^* \cdot f_{max}}, (5)$$

where the frequencies $f_i^*$ are calculated in the same way as given in Eq. (3).

The eight values of $\sigma_{Ni}^2$ can be calculated and stored in a look-up table in the microcontroller. As a result, the second method does not require significantly more resources than the first one. In Fig. 2 we compare the performance of the two methods in case of $1/f$ noise. We can observe, as expected, that the weighted method gives exactly the $1/f$ spectrum. Note that the direct method also gives a good approximation of the spectrum. By multiplying the PSD values with the frequency we can enlarge the deviation from the ideal $1/f$ spectrum. In the case of the first frequency band of the direct method, the deviation is significant. Thus, we can conclude that the results of the simulations are in accordance with the theoretical results.

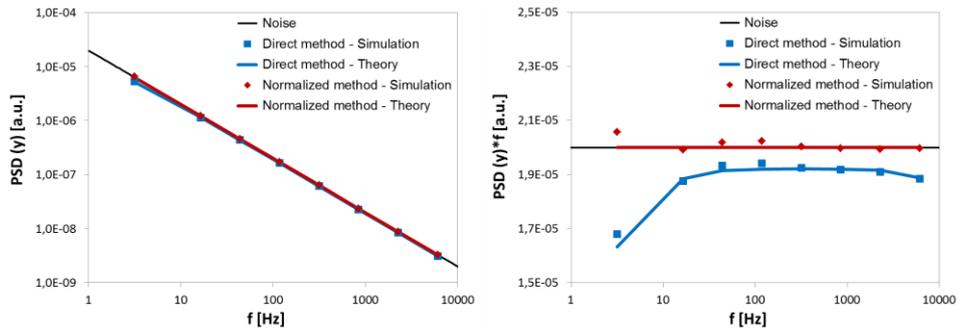

Fig. 2. Comparing the result of the presented methods in the case of $1/f$ noise. On the right side the PSD is multiplied by the frequency to enlarge the deviations. Note that the theoretical values of the reconstructed PSD are interpreted only in eight points, we connected them only to increase the readability of the figure.

On Fig. 3 we examine the performance of the two methods in case of $1/f^\alpha$ noises. While the normalized method does not give the exact PSD values, it still performs much better than the direct method. For this reason, in the following sections of the paper we will only use the normalized method.

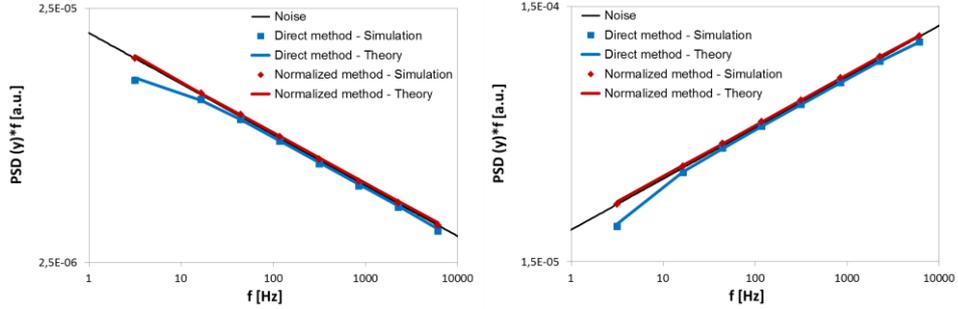

Fig. 3. Comparing the performance of the two methods in the case of $1/f^{1.2}$ (left) and $1/f^{0.8}$ (right). The vertical scale is logarithmic.

## Evaluation of the required number of filters

In order to test the resolution of the spectral reconstruction method we added a plateau (or peak) to $1/f$ noise PSD as defined by Eq. (6) and illustrated on Fig. 4. The maximum deviation from the $1/f$ noise occurs at $f_c$, the center frequency of the plateau (316 Hz), the width of the peak is $\Delta f$ (158 Hz) and the amplitude is two times higher than the amplitude of $1/f$ noise at the center frequency ($A = 1$).

$$S(f) = \frac{1 \cdot V^2 Hz}{f \cdot f_{max}} \cdot \left(1 + A \cdot \exp\frac{(f - f_c)^2}{\Delta f^2}\right), \qquad (6)$$

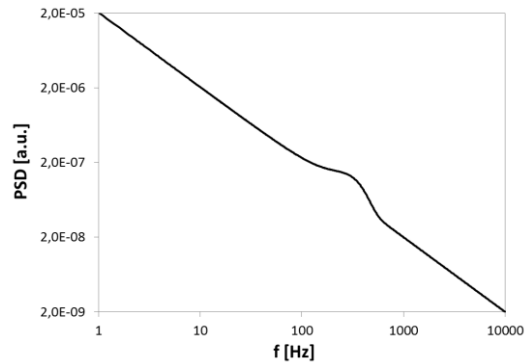

Fig. 4. The PSD of the noise used to test the resolution of the spectral reconstruction method.

On the left side of Fig. 5 we can observe the result of the reconstruction using eight filters. We can see that the peak in the reconstruction is significantly wider than the original one due to the moderate roll-off of the low pass filters. In the case of sixteen filters, the width does not decrease, however, we get the wide peak with much higher resolution.

The slow decay of first order filters (20 dB/decade) is the major limiting factor of the resolution of the method. To increase the resolution one may use higher order filters, deconvolution based methods or even more complex algorithms [18]. However, these methods would significantly increase the complexity and power consumption of the sensor node.

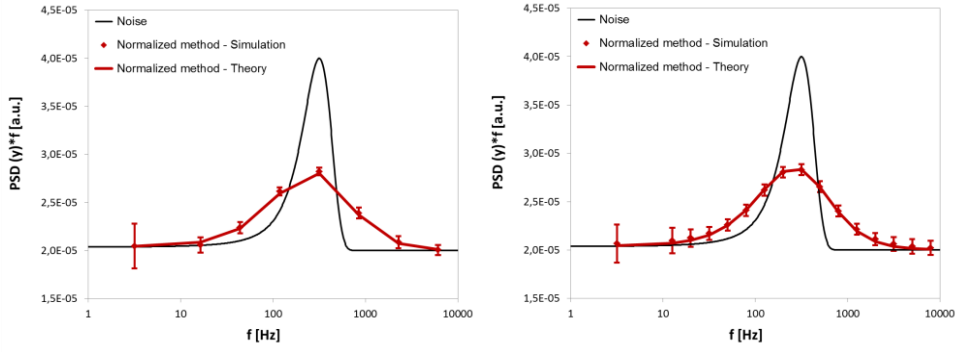

Fig. 5. The reconstruction of a plateau (peak) in the PSD. Both the PSD and the reconstruction is multiplied by f to increase the readability of the figure. On the left 8 filters were used, while on the right 16 filters.

# Results and discussion

In the next section we analyze the performance of the proposed $1/f$ normalized PSD reconstruction method. We perform test using $1/f^\alpha$ noises with different exponents, test the effect of different peak heights and widths and we also test our method by simulating signals with PSD identical of different published measurement results.

## Reconstructing artificial noises with different PSD-s

On the left side of Fig. 6 we demonstrate the performance of the normalized method in the case of different $1/f^\alpha$ noises. The exponents of the noises were 0.8, 0.9, 1.0, 1.1, and 1.2 respectively. The error bars were calculated based on 20 simulations each, showing the standard deviation of the variance. In the case of the first band, the error is higher than in the case of higher frequency bands, but still very low. This low error value can give us the opportunity to reduce the length of the measurement and to save power.

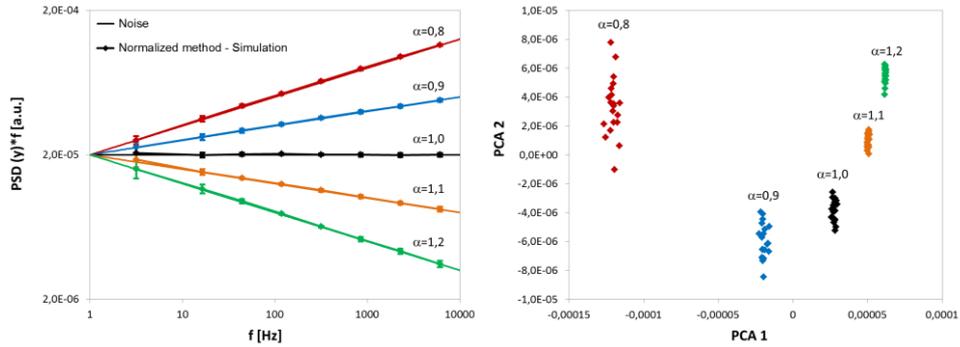

Fig. 6. On the left the reconstruction of different $1/f^\alpha$ noises is illustrated. On the right the result of the PCA analysis for the same noises is shown.

In our previous work [5-7] we have used PCA based pattern recognition methods in order to distinguish between different kinds of gases. In the referred works the PCA was calculated from the averaged spectrums, each of them consisting of around 2000 points. Our reconstruction method provides only 8 points as an input for the PCA analysis. However, as seen on the right side of Fig. 6, these 8 points give enough information to distinguish between different noise exponents while the processing time is reduced radically.

In addition, we have examined the effect of amplitude of the peak. The parameters *A* for the peaks were −0.5, 0.0, 0.5, 1.0, and 2.0 respectively. As can be seen in Fig. 7 the different PSD-s can be clearly distinguished from each other both by spectral recognition, and by PCA analysis.

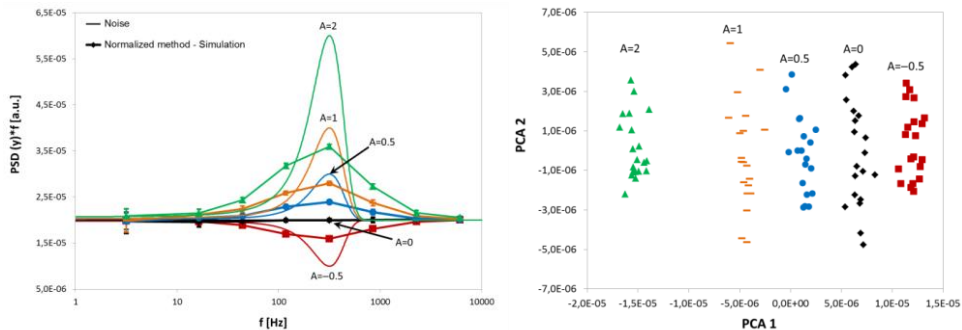

Fig. 7. The spectral reconstruction in the case of different PSD peak amplitudes.

Moreover, we have tested the reconstruction algorithm for peaks with different widths. $\Delta f$ values of 15 Hz, 30 Hz, 70 Hz, 157 Hz and 300 Hz were used. As it is displayed on Fig. 8, wide peaks with high power can be rather accurately recognized. In the case of narrow peaks, the frequency bands close to them differ from the reference values; this difference is slightly higher than the SD of the variance. However the PCA point groups in the case of narrow peaks overlap (Fig. 9). As a result, we conclude that PCA analysis in its current form may not be the best classification method for the current application. At the same time, the points of the reconstructed PSD are suitable as an input data for more elaborated pattern recognition methods like support vector machine [15, 19, 20].

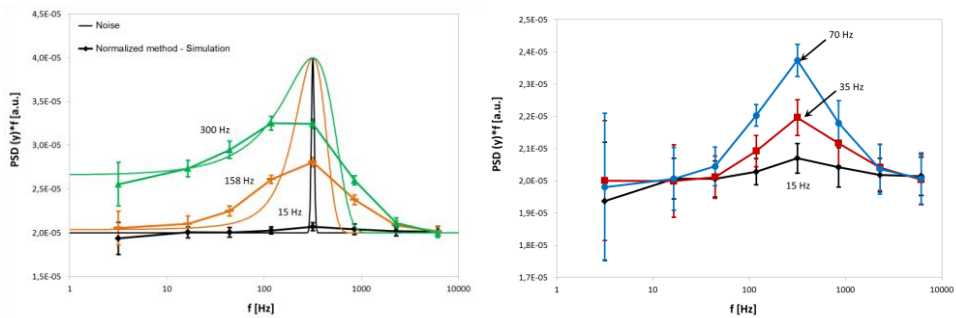

Fig. 8. The effect of the bandwidth of the peaks.

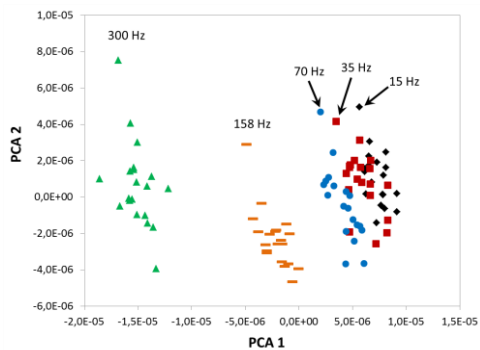

Fig. 9. The PCA analysis of the effect of the bandwidth of the peaks.

## Reconstruction examples of measured spectra

In order to test our algorithm, we have generated noises based on the PSD of noise depending on bacterial odors [14 Fig. 6]. As illustrated in Fig. 10 the reconstruction can detect the presence of bacteria and distinguish between different bacterium types. The groups of dots can also be distinguished in the calculated PCA plots.

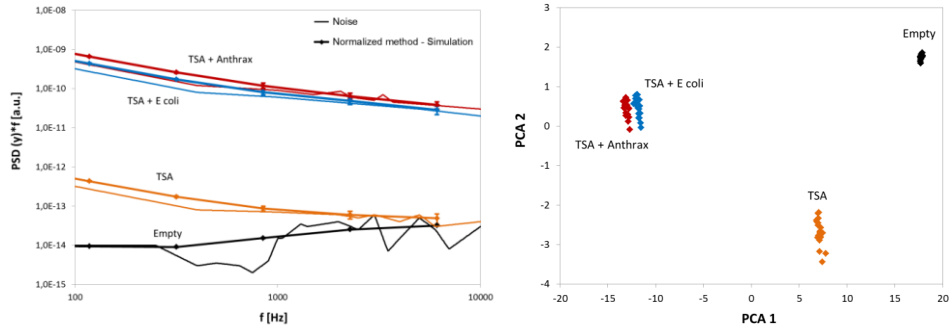

Fig. 10. Reconstruction of PSD in the case of bacteria odor sensing on the left, and the results of the PCA algorithm on the right.

In the second analysis we generated the test noise with similar PSD that had been previously observed in a carbon nanotube gas sensors in 50 ppm CO and 50 ppm $N_2O$ environments. Our goal was to contrast the performance of the gas detection compared to the PCA based method described in [5]. For each measurement point we calculated the PSD of the noise by averaging 100 spectra consisting of 4096 samples per spectrum. The sampling frequency was 10 kHz and we used the spectrum components over 100 Hz for the PCA analysis. The result of the analysis for the simulated signals is shown on the left side of Fig. 11. Our method uses the same measurement length (40.96 s) while the sampling frequency was 1 kHz per filter. As it can be seen on the right side of Fig. 11, our method performs as well as the original one. Note that while the value of PCA components cannot be directly compared to each other since they used input data with different orders of magnitude, the signal-to-noise ratio can be compared based on the results.

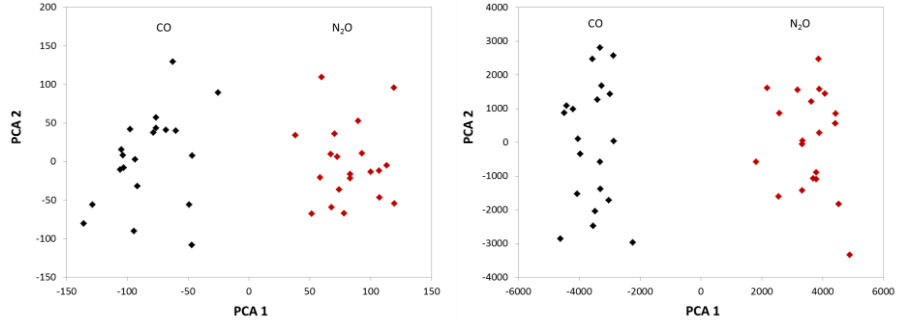

Fig. 11. Comparing the performance of pattern recognition method introduced in [5] (left side) with our PSD reconstruction method (right side) in the case of simulated noises corresponding to a carbon nanotube sensor in 50 ppm CO and 50 ppm $N_2O$ environments.

# Conclusion

In the present paper we investigated two spectral reconstruction methods that use output signals of an analogue low-pass filter bank. The reconstruction methods were designed to be used with our recently presented low power wireless sensor node and optimized to estimate PSDs similar to the PSD of $1/f$ noises. We tested the performance of the methods using different spectral densities and found that – although the resolution is rather limited – even slightly different spectral densities can be distinguished. This feature supports low power FES measurements in small devices. Moreover, the output data of the PSD reconstruction can be used as an input for different pattern recognition algorithms. We compared the performance to previously used pattern recognition methods, and we found that the same precision requires at least 50 times less operations and 20 times less memory size. Since our algorithm can be implemented by using only 16 or 32 bit integer arithmetic, combined with the low memory needs, it can be embedded into ultra-low-power microcontrollers, while the FFT based method exceeds the possibilities of the hardware.

The $1/f$ normalized method is fine-tuned for processing $1/f$ like noises, but by modifying Eq. (4) and (5) it can be applied for other signal types as well. The possible applications include gas mixture detection, bacterial odor and scent sensing by very low power intelligent sensor nodes. Note that reliability testing, vibration monitoring or other tasks requiring spectral analysis done by power efficient compact devices might also be supported.

# Acknowledgements

This research was supported by the European Union and the State of Hungary, co-financed by the European Social Fund in the framework of TÁMOP-4.2.4.A/2-11/1-2012-0001 'National Excellence Program'.